\documentclass[twocolumn,showpacs,superscriptaddress,preprintnumbers,amsmath,amssymb]{revtex4}

\usepackage{graphicx}
\usepackage{dcolumn}
\usepackage{bm}
\usepackage{hyperref}

\newcommand{\bq}{\begin{equation}}
\newcommand{\eq}{\end{equation}}
\newcommand{\bqa}{\begin{eqnarray}}
\newcommand{\eqa}{\end{eqnarray}}
\newcommand{\nn}{\nonumber \\}

\def\be     {\begin{equation}}
\def\ee     {\end{equation}}
\def\bea        {\begin{eqnarray}}
\def\eea        {\end{eqnarray}}
\def\bnn    {\begin{eqnarray*}}
\def\enn    {\end{eqnarray*}}

\begin{document}

\title{A criterion for the nature of the superconducting
transition \\in strongly interacting field theories : Holographic
approach}

\author{ Ki-Seok Kim$^{1,2}$, Kyung Kiu Kim$^3$,
Youngman Kim$^{1,2}$ and Yumi Ko}

\affiliation{Asia Pacific Center for Theoretical Physics,
Pohang, Gyeongbuk 790-784, Republic of Korea \\ $^2$ Department of
Physics, Pohang University of Science and Technology, Pohang,
Gyeongbuk 790-784, Korea \\ $^3$ Institute for the Early Universe,
Ewha Womans University, Seoul 120-750, KOREA}



\begin{abstract}
It is beyond the present techniques based on perturbation theory to
reveal the nature of phase transitions in strongly interacting field
theories. Recently, the holographic approach has provided us with an effective
dual description, mapping strongly coupled conformal field
theories to classical gravity theories. Resorting to the
holographic superconductor model, we propose a general criterion
for the nature of the superconducting phase transition based on
effective interactions between vortices. We find ``tricritical''
points in terms of the chemical potential for U(1) charges and an
effective Ginzburg-Landau parameter, where vortices do not
interact to separate the second order (repulsive) from the first
order (attractive) transitions. We interpret the first order transition as the
Coleman-Weinberg mechanism, arguing that it is relevant to
superconducting instabilities around quantum criticality.
\end{abstract}


\maketitle

Interactions between vortices contain information on the nature of
the superconducting transition. They change from repulsive to
attractive, decreasing the Ginzburg-Landau parameter $\kappa$, the
ratio between the penetration depth of an electromagnetic field
and the Cooper-pair coherence length
\cite{Vortex_Interaction1,Vortex_Interaction2,Vortex_Interaction3}.
Combined with either the $\epsilon = 4 - d$ expansion or the $1/N$
approximation in the Abelian-Higgs model \cite{FIFT,Review_1vs2},
one finds that the noninteracting point for vortices at $\kappa =
\kappa_{t}$ ($\sim 1/\sqrt{2}$) is identified with the tricritical
point, where the nature of the superconducting transition changes
from  second order ($\kappa > \kappa_{t}$) to  first order
($\kappa < \kappa_{t}$) \cite{Review_1vs2}. Quantum corrections
due to electromagnetic fluctuations are the mechanism, referred as
the fluctuation-induced first-order transition \cite{FIFT} or
Coleman-Weinberg mechanism \cite{CWM}.

The situation is much more complicated when correlated electrons
are introduced. In particular, superconducting instabilities are
ubiquitous in the vicinity of quantum critical points
\cite{QCP_Review1}, where quantum critical normal states are often
described by strongly interacting conformal field theories.
Although one can integrate over such interacting fermions, the
resulting effective field theory contains a lot of singularly
corrected terms for Higgs fields, which originate from quantum
corrections due to abundant soft modes of particle-hole and
particle-particle excitations near the Fermi surface
\cite{QCP_Review2,FMQCP}. Furthermore, the Fermi surface problem
turns out to be out of control \cite{SSL,Max} since not only
self-energy corrections but also vertex corrections should be
introduced self-consistently. It is far from reliability to
evaluate effective interactions between vortices in this problem.

Recently, it has been clarified that strongly coupled conformal
field theories in $d$-dimension can be mapped into classical
gravity theories on anti-de Sitter space in $d+1$-dimension
(AdS$_{d+1}$) \cite{AdS_CFT_Conjecture1,AdS_CFT_Conjecture2}. This
framework has been developed in the context of string theory,
refereed as the AdS/CFT correspondence. See Ref.
\cite{AdS_CFT_Review} for a review. Immediately, it has been
applied to various problems beyond techniques of field theories:
non-perturbative phenomena in quantum chromodynamics (AdS/QCD or
holographic QCD) \cite{hQCD_review}, non-Fermi liquid transport
near quantum criticality
\cite{AdS_CMP_TR1,AdS_CMP_TR2,AdS_CMP_TR3} and superconductors
\cite{H sconductor gubser,H superconductor HHH} in condensed
matter physics (AdS/CMP), and etc.

In this letter we propose a general criterion for the first-order
superconducting transition based on the holographic approach. We
take the holographic superconductor model \cite{H superconductor
HHH} as an effective low-energy model in the dual description for
certain classes of strongly interacting field theories. The
asymptotic vortex solution \cite{H sconductor vortex} turns out to
play a central role in the nature of the superconducting
transition. We suggest ``tricritical'' points in terms of the
chemical potential for U(1) charges and an effective
Ginzburg-Landau parameter, where vortices do not interact to
separate the second order (repulsive) from the first order
(attractive). We interpret the first-order transition as the
Coleman-Weinberg mechanism \cite{FIFT,CWM}, arguing to be relevant
to superconducting instability around quantum criticality.

We start from the holographic superconductor model in  AdS$_{4}$ with radius $L$
\begin{eqnarray} \label{Action}
S \!&=\!& \frac{1}{2 \kappa_p^2} \int\! d^4 x \sqrt{-g} \,\Bigl[ R +\frac{L^2}{6} -\frac{L^2}{4}F^2 -\frac{1}{2}(\mathcal{D} \eta)^2 \nn
&&- \frac{1}{2} \eta^2 (e A_\mu-\mathcal{D}_\mu \phi)^2 -
\frac{m^2}{2}\eta^2  \Bigr] ,
\end{eqnarray}
where the complex scalar field is decomposed into the amplitude
$\eta$ and the phase $\phi$, and $A_{\mu}$ is the bulk gauge
potential with the field strength $F =dA$.
$\kappa_{p}$ is the Planck's constant. In this work we set
$e=1$ and $m^2=-2/L^2$ and consider the probe limit. The
background metric is given by
\begin{eqnarray}
ds^2 = \frac{L^2}{z^2}\Bigl( -\alpha^2 f(z) dt^2 + dx_1^2 + dx_2^2
+ \frac{dz^2}{ f(z)} \,\Bigr),
\end{eqnarray}
with $f(z)= 1-z^3$. The Hawking temperature is given by $T =
\frac{3\alpha}{4 \pi}$. 

Equations of motion read
\begin{eqnarray}\label{eom}
&& {\mathcal{D}}^2 \eta -m^2\eta - \eta \,Q_\mu^2  = 0 , \nn && L^2 \,{\mathcal{D}}_\mu
B^{\mu\nu} - \eta^2 Q^\nu = -L^2 \,{\mathcal{D}}_\mu X^{\mu\nu} ,
\end{eqnarray}
where $Q_\mu \equiv A_\mu -{\mathcal{D}}_\mu \phi$ is the gauge
invariant superfluid four-velocity and $B_{\mu\nu}$ is its field
strength. $X_{\mu\nu}$ is $\partial_{[\mu}\partial_{\nu]} \phi$
which can be replaced with delta functions for centers of
vortices.

Now we calculate effective interactions between vortices. The
effective interaction will be determined by the change of a single
vortex solution in a widely separated vortex-lattice configuration
with a lattice spacing $d_L$
\cite{Vortex_Interaction1,Vortex_Interaction2}. The variation of
the single vortex solution occurs dominantly around the boundary
of two vortices $\sim d_L/2$, proven to coincide with an asymptotic
solution of the single vortex. In this respect we proceed as
follows. First, we find the asymptotic solution of a single vortex
away from the vortex core. Second, we show that the variation of
the vortex solution is given by the asymptotic solution. Third, we
represent the vortex interaction in terms of this solution.

We introduce the following ansatz for an
asymptotic solution of the single vortex configuration
\begin{eqnarray}\label{separable ansatz}
&& \!\!\!\!\! \!\!  s(\vec x,z) \equiv {\eta^v}(r,z) -{\eta}^{s}(z) = A(z)\,
R(r)\, , \nn
&&\!\!\!\!\! \!\! q_t(\vec x ,z) \equiv{Q_{\,t}^v}\,(r,z) - { Q}_{\,t}^{s}(z) = B(z)\, R(r) \,,\nn
&&\!\!\!\!\! \!\!  q_i(\vec x ,z) \equiv {Q}_{\,i}^{v}(r,z) - {
Q}_{\,i}^{s}(z) = C(z)\, \mathcal{R}_i ( r ) \,,
\end{eqnarray}
where the superscript $v$ represents the single vortex solution
and $s$ denotes the uniform solution with the radial coordinate
$r$ or rectangular coordinates $x_i$ in two dimension. Then,
Eq.~(\ref{eom}) becomes
\begin{eqnarray}
&&\!\!\!\!\! \!\! \left(\nabla^2 - \kappa_1^2 \right) R(\vec {
x})=0 , \quad \left(\nabla^2 -  \kappa_2^2
\right)\mathcal{R}_i(\vec {x})=0, \nn &&\!\!\!\!\! \!\! \!A''\!\!
+\! \Bigl(\frac{f'}{f} -\frac{2}{z}\Bigr) A'\!\! + \Bigl(
\frac{\kappa_1^2}{f}+ \frac{2 }{z^2 f}+ \frac{ \,{Q_{t}^{s}}^2}{
f^2} \Bigr) A + \!\frac{2 \eta^{s} Q_{t}^{s}}{ f^2}B = 0 , \nn &&
\!\!\!\!\! \!\!B''\!\! + \Bigl( \frac{\kappa_1^2}{ f}-\frac{
{\eta^s}^2}{z^2 f}\Bigr) B - 2 \frac{{ \eta}^{s}{ Q}_{t}^{s}}{z^2
f} A = 0 , \nn &&\!\!\!\!\! \!\! C'' + \frac{f'}{f}\, C'+\Bigl(
\frac{\kappa_2^2}{f} - \frac{{\eta^s}^2}{z^2 f}\Bigr) C = 0,\,
\label{asymp eom z}
\end{eqnarray} in  $q_{z}=0$  gauge. It is straightforward to see
$R(r) = K_0(\kappa_1 r)$, $\mathcal{R}_\theta(r) = K_1(\kappa_2
r)$, and $\mathcal{R}_r(r) = 0$ in polar coordinates. Here,
$\kappa_{1}$ and $\kappa_{2}$ are constants for separation of
equations. Scaling the radial coordinate by $\bar r = \kappa_2 r$,
we find that Eq.~(\ref{asymp eom z}) can be rewritten in terms of
only a single parameter $\kappa = \frac{1}{\sqrt{2}}
\frac{\kappa_{1}}{\kappa_{2}}$. For other equations, we need to
solve them numerically, taking the regularity conditions at the
horizon. It turns out that resulting solutions depend on $\kappa$ and
 $a$, $b$, $c$, which are defined at the horizon, $a=A(1)$, $b=B'(1)$, and $c=C(1)$. In
addition, we find that such solutions are characterized only by
$b/a$ and $\kappa$ due to the scaling symmetry of Eq.~(\ref{asymp
eom z}). See appendix A for the numerical analysis
\cite{Supplementary}.

Having the asymptotic solution, we evaluate the effective
interaction between vortices in the dilute vortex-lattice
configuration \cite{Vortex_Interaction1,Vortex_Interaction2}. We
introduce
\begin{eqnarray} \label{expand fields}
 &&\!\!\!\!\!\eta = \eta^{v} +
\delta \eta\, , \quad Q_t = Q_t^{v} + \delta Q_t \,, \nn
 &&\!\!\!\!\! Q_i =
Q_i^{v} + \delta Q_i \,, \quad \phi = n\theta + \delta \phi \,,
\end{eqnarray}
where the solution with the superscript $v$ represents the single
vortex configuration in a Wigner-Seitz cell, while the ``$\delta
$" part expresses the variation of the single vortex configuration
around the boundary of the Wigner-Seitz cell. $n$ is the winding
number of the vortex. $\delta \phi = \sum_{\hat i\neq 0} n
\arg(\vec x-\vec x_{\hat i})$ is chosen for a multi-vortex
configuration, where $\vec x_{\hat i}$ is the core position of
each vortex. $\delta \eta$ and $\delta Q_t$ would be much smaller
than $\eta^{v}$ and $Q_t^{v}$ inside the Wigner-Seitz cell,
respectively. On the other hand, it is not obvious if $\delta Q_i$
is much smaller than $Q_i^{v}$ near the boundary of the
Wigner-Seitz cell because $Q_i^{v}$ will be also small. However,
it is natural to expect that $\delta \eta$ and $\delta Q_t$ are
much larger than $\delta Q_i^2$ near the boundary
\cite{Vortex_Interaction1}. As a result, we obtain the following
linearized equations of motion near the boundary
\begin{eqnarray}
&&\!\!\!\!\! {\delta \eta}''\! + \Bigl(\frac{f'}{f}
-\frac{2}{z}\Bigr){\delta  \eta}' + \Bigl(
\frac{{\nabla}^2-{{ Q}_i^v}^2}{f}+ \frac{2 }{z^2 f}+\frac{\,
{Q_t^v}^2}{ f^2}  \Bigl) \delta
 \eta \nn
 && \!\!\!\!\!+ 2\frac{ \eta^{v}  Q_t^{v}
}{f^2} \delta  Q_t -2 \frac{ \eta^{v}
Q^{v}_i}{f} \delta  Q_i =0 , \nn
&&\!\!\! \!\!{{\delta
Q}_t}\!''\! + \Bigl( \frac{{\nabla}^2}{f}-\frac{
{\eta^v}^2}{z^2 f}\Bigr) {\delta  Q}_{t} -2\frac{
\eta^{v}  Q_t^{v}}{z^2 f} \delta  \eta = 0 , \nn
&&\!\!\!\!\! {\delta Q}_{i}''\! + \frac{f'}{f}\, {\delta
Q}_{i}'+\Bigl( \frac{{\nabla}^2}{f} - \frac{{
\eta^v}^2}{z^2 f}\Bigr) {\delta Q}_{i} - 2\frac{{
\eta}^{v} Q_i^{v}}{z^2 f} \delta  \eta=0.
\label{linearlinear eom}
\end{eqnarray}
An important aspect is that these equations are essentially the
same with those for the asymptotic configuration of the single
vortex, valid when $\eta^{v} \gg \delta \eta$ and $Q_t^{v} \gg
\delta Q_t$ with $\eta^{v}$ or $Q_t^{v}  \gg \delta Q_i^2$. This
property leads us to write down the variation of the solution in
terms of the asymptotic solution for the single vortex
configuration
\begin{eqnarray}\label{asymp sol}
\delta  \eta =  \sum_{\hat i \neq 0} {s}(\vec
x - \vec x_{\hat i}, z) ,  \quad\delta  Q_\mu =  \sum_{\hat i\neq 0}
{q}_\mu(\vec x - \vec x_{\hat i},z) .
\end{eqnarray}

Expanding the action (\ref{Action}) around a vortex solution to second order and using equations of motion (\ref{eom}), we arrive at
\begin{eqnarray}
\!\!\!\!\!\!\delta \Omega^{(2)}\!\!\!&=&\!\!\frac{1}{T} \int_{\hat
i} dz d^2 x \partial_\mu \Bigl\{\sqrt{-g}\,\Bigl[\, \delta\eta
\mathcal{D}^\mu \Bigl(  \eta^{v} + \frac{1}{2}\delta \eta \Bigr)
\nonumber\\&&~~~~~~~~~~~~~~~~~~~~+\delta Q_\nu \Bigl(
{B^v}^{\mu\nu}+ \frac{1}{2}\delta B^{\mu\nu}  \Bigr) \Bigr]\Bigr\}
.
\end{eqnarray}
We observe that only surface terms contribute to the correction
for the grand potential, where these boundaries correspond to the
AdS$_{4}$ boundary ($z=0$), the horizon ($z=1$), and the boundary
of the Wigner-Seitz cell. The regularity condition on the horizon
does not allow contributions from the horizon. In addition, the
dilute vortex configuration guarantees that the contribution along
the AdS$_{4}$ boundary is much smaller than that from the boundary
of the Wigner-Seitz cell \cite{Supplementary}. Therefore, the
relevant contribution is from the correction at the boundary of
the Wigner-Seitz cell, which is attainable by the asymptotic
solution (\ref{asymp sol}). Finally, we obtain the change of the
grand potential for the $i^{th}$ cell,
\begin{eqnarray}\label{interaction 2}
\delta \Omega^{(2)}
\!\!\!&\sim&\!\!\! \frac{\alpha }{ T } \int_0^1\!\! dz \oint_{\hat
i} d  l ~ \hat n \cdot \Bigl\{ \frac{1}{  z^2} \delta
\eta  \nabla ( \eta^v + \frac{1}{2}\delta  \eta)
\nn
\!\!\!&&\!\!\! -\frac{1}{f(z)} \delta  Q_t  \nabla (
Q^v_t + \frac{1}{2} \delta Q_t ) + \delta \vec{Q}
\times \!\nabla\!\times ({\vec{Q}}^v + \frac{1}{2} \delta
\vec{Q}) \Bigr\} \nn &=&\!\!\! \frac{4 \pi^2 }{3} \!\sum_{\hat i
\neq 0} \left[  \mathcal C K_0(   \bar r_{\hat i}) -
(\mathcal A - \mathcal B) K_0( \sqrt{2}\kappa \bar
r_{\hat i}) \right]\!
\end{eqnarray}
with $\mathcal A = \!\!\int_0^1\!\! dz \frac{ A(z)^2}{ z^2}$,
$\mathcal B =\!\! \int_0^1\!\! dz \frac{ B(z)^2}{f(z)}$ and
$\mathcal C =\!\! \int_0^1\!\! dz  C(z)^2$. For the analytic
expression in the last line of Eq.~(\ref{interaction 2}), we used
an identity in Ref. \cite{identity}.

It is possible to understand the physical meaning of
Eq.~(\ref{interaction 2}). The interaction potential consists of
both first order and second order contributions in ``$\delta$",
where the former represents interactions between the $i^{th} = 0$
vortex and others $i^{th} \not= 0$, and the latter expresses those
between other vortices $i^{th} \not= 0$ except for the $i^{th} =
0$ vortex.

This expression is formally identical to the effective interaction
between vortices in the Abelian-Higgs model, where the first term
results from the variation of the supercurrent while the second
originates from that of the Higgs field around the boundary
\cite{Vortex_Interaction1,Vortex_Interaction2}. An important
ingredient is that coefficients of the vortex interaction are
given by integrals in the $z$-direction. In addition, the $\kappa$
dependence of the interaction potential is much more complicated
since such coefficients are functions of the parameter $\kappa$.
In this respect the role of the parameter $\kappa$ is not
completely clear yet although tuning $\kappa$ results in the
change of the vortex interaction.

\begin{figure}[t]
\center {\includegraphics[width=2.7cm]{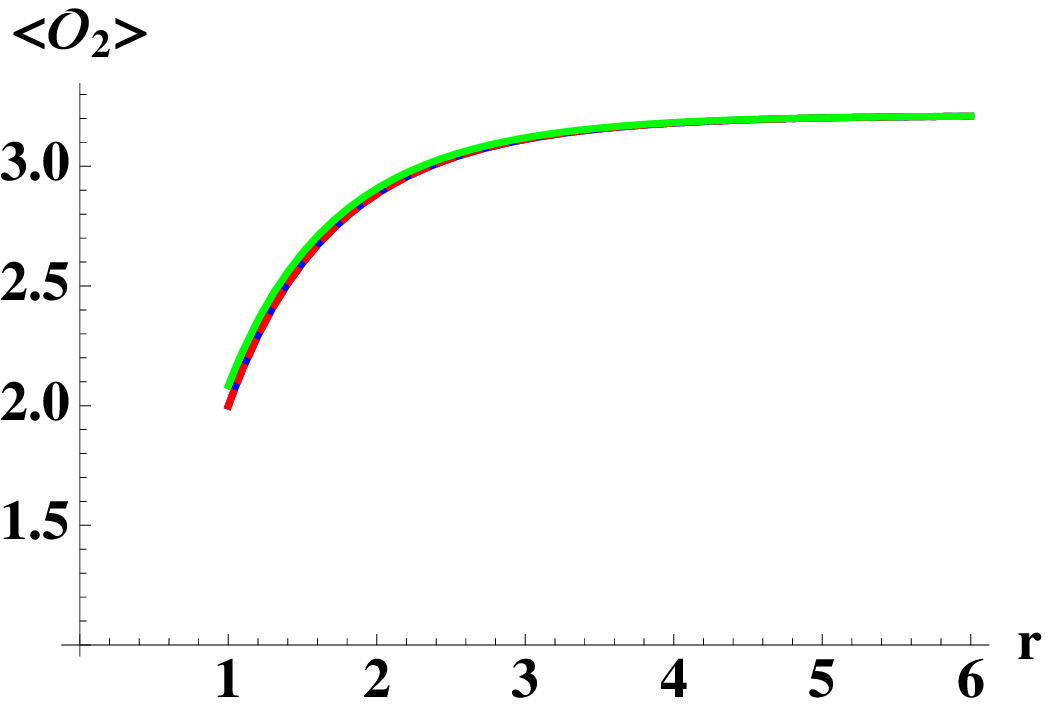}}
{\includegraphics[width=2.7cm]{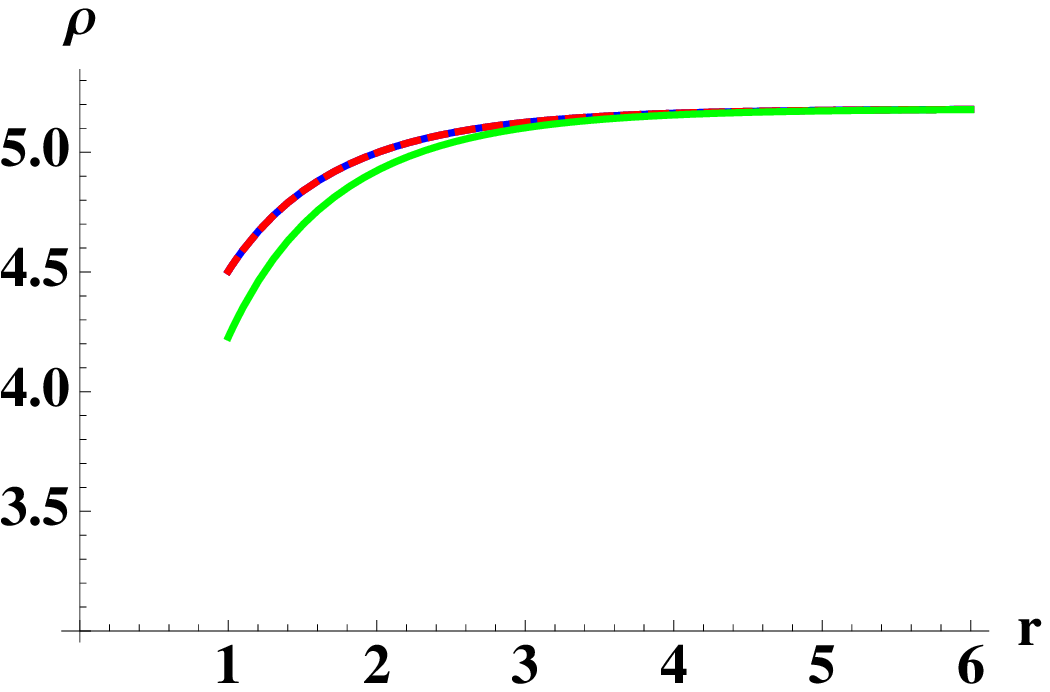}}
{\includegraphics[width=2.7cm]{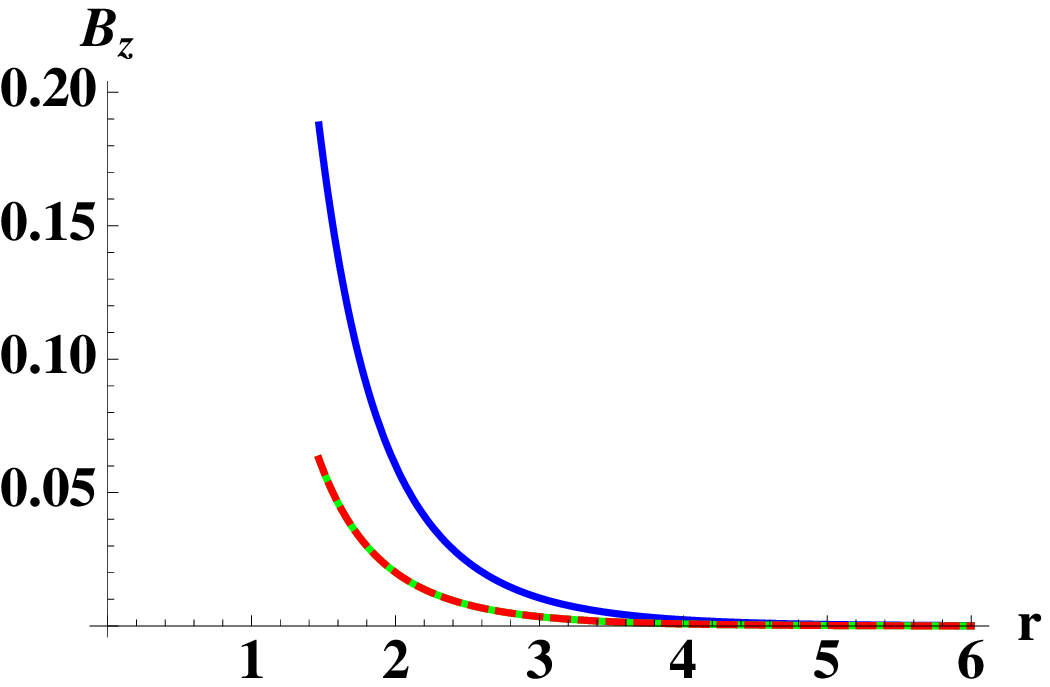}} \caption{Asymptotic vortex
solutions for dimension 2 condensation, charge density, and
magnetic flux, respectively, where $\kappa=1$ is used. Both the blue line with  $[\,a/b=-1$, $a/c=2\,]$ and the green line with
$[\,a/b=1$, $a/c=6\,]$ correspond to repulsive interactions between
vortices while the dashed red line with $[\,a/b=-1$, $a/c=6\,]$ results
in attractive interactions.} \label{fig1}
\end{figure}

Figure \ref{fig1} shows dimension 2 condensation, charge density,
and magnetic flux for the asymptotic single-vortex configuration,
respectively. It is interesting to observe that when U(1) charge
density decreases rapidly near the vortex core, the effective
interaction between vortices becomes more repulsive.
As long as $b/a$ remains positive, we do not see any change from
repulsive to attractive interactions. In this case the system
might lie in a deep type II regime. Therefore, we focus on $b/a <
0$ hereafter to study the superconducting transition.

\begin{figure}[t]
\center {\includegraphics[width=3.2cm]{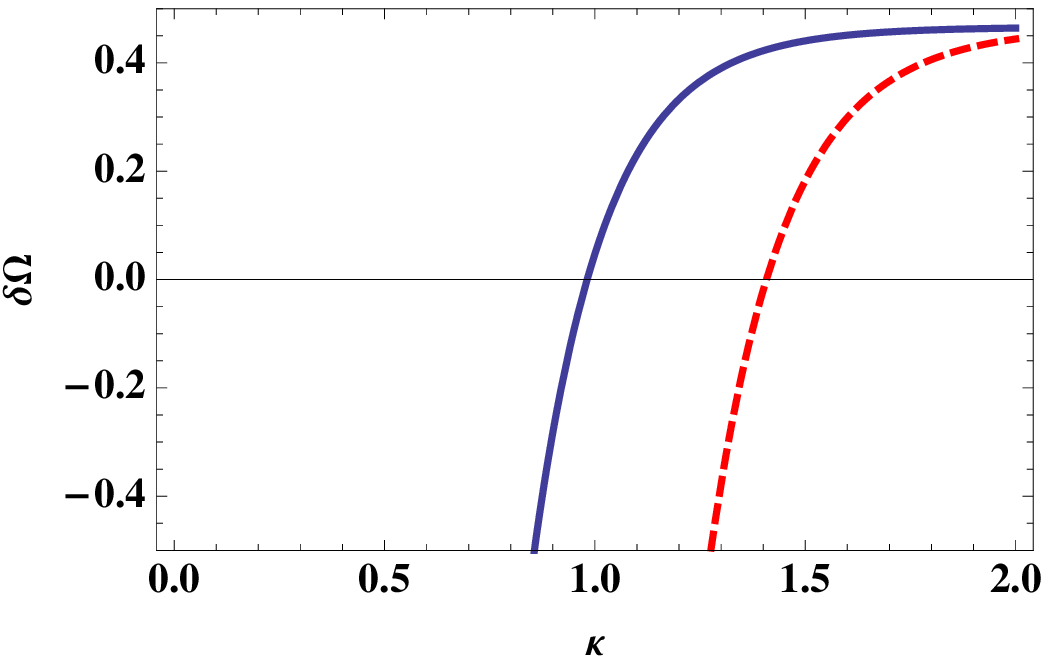}}\quad
{\includegraphics[width=3.2cm]{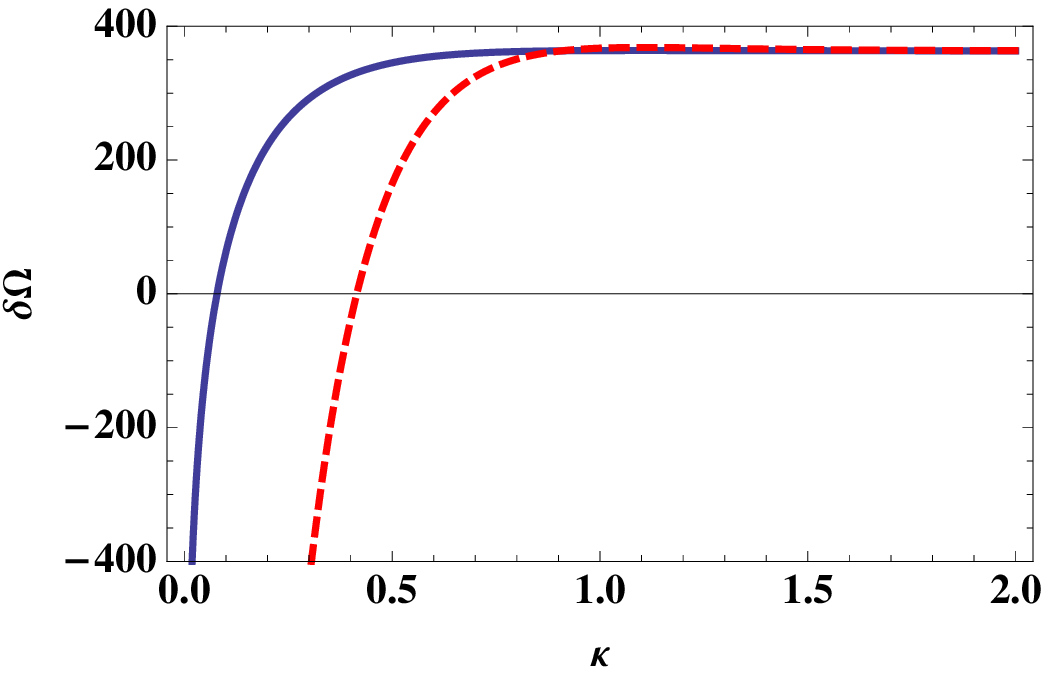}} \caption{Effective
interactions between vortices as a function of $\kappa$.
With decreasing $\kappa$ the effective potential changes from
repulsive to attractive. The nearest interaction was considered where $d_L=3$. Left: $T/T_c = 0.907$, $b/a = -1$, $ a/c=
3$ (Blue Solid), and $a/c= 10$ (Red Dashed). Right: $T/T_c =
0.388$, $b/a = -1$, $ a/c= 3$ (Blue Solid), and $a/c= 10$
(Red Dashed).}\label{fig2}
\end{figure}

We classify our systems into four classes under the condition of
$b/a<0$, depending on the density of U(1) charges and the ratio of
$a/c$. First, we fix the density of U(1) charges, determining the
chemical potential. We expect that the regime with $a/c \ll 1$
belongs to the type II superconductivity because the first term of
the vortex interaction in Eq.~(\ref{interaction 2}) becomes larger
than the second term, resulting in repulsive interactions.
Physically, this relation implies strong supercurrents around the
vortex core, consistent with the picture of type II. On the other
hand, the regime with $a/c \gg 1$ will show that interactions
between vortices change from repulsive when $\kappa \gg 1$ to
attractive when $\kappa \ll 1$. Notice that $\kappa$ is introduced
into the second term, reducing it with $\kappa \gg 1$ and
enhancing it with $\kappa \ll 1$. Both $\mathcal A$ and $\mathcal
B$ are positive definite, decreasing monotonically as we increase
$\kappa$. We uncover that the regime with $b/a < 0$ gives rise to
$\mathcal A - \mathcal B > 0$, allowing the possibility for the
change of interactions. Fig.~\ref{fig2} confirms our expectation,
that is, interactions between vortices become attractive when
$\kappa < \kappa_{t}$, where $\kappa_{t}$ can be regarded as the
tricritical point.

\begin{figure}[t]
\center {\includegraphics[width=4.2cm]{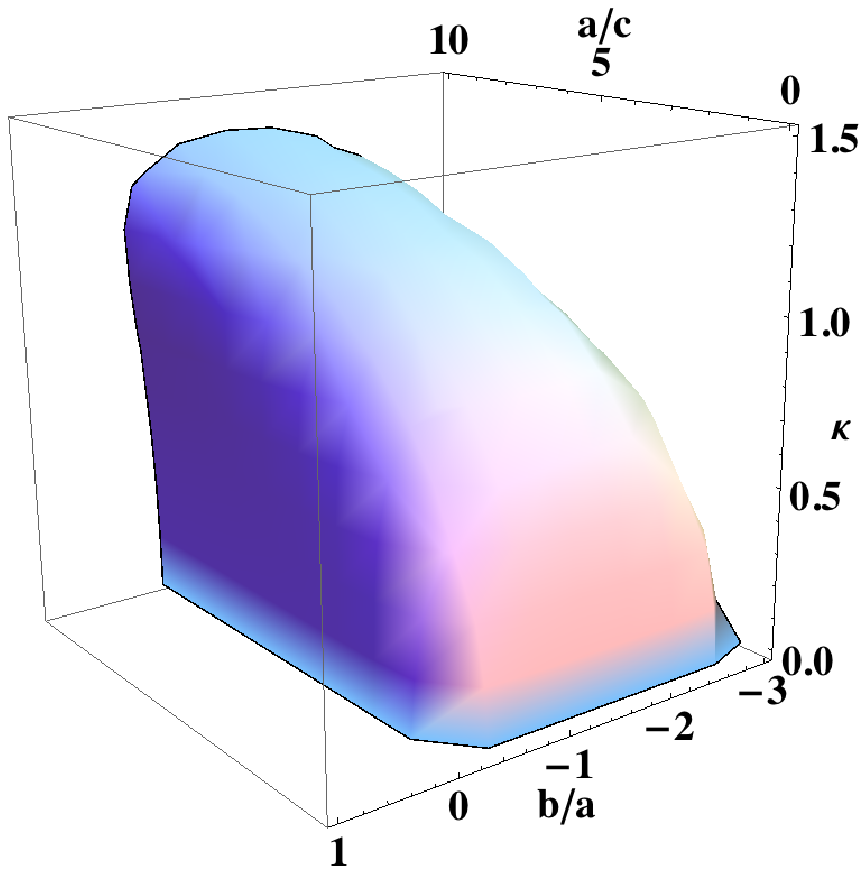}}
{\includegraphics[width=4cm]{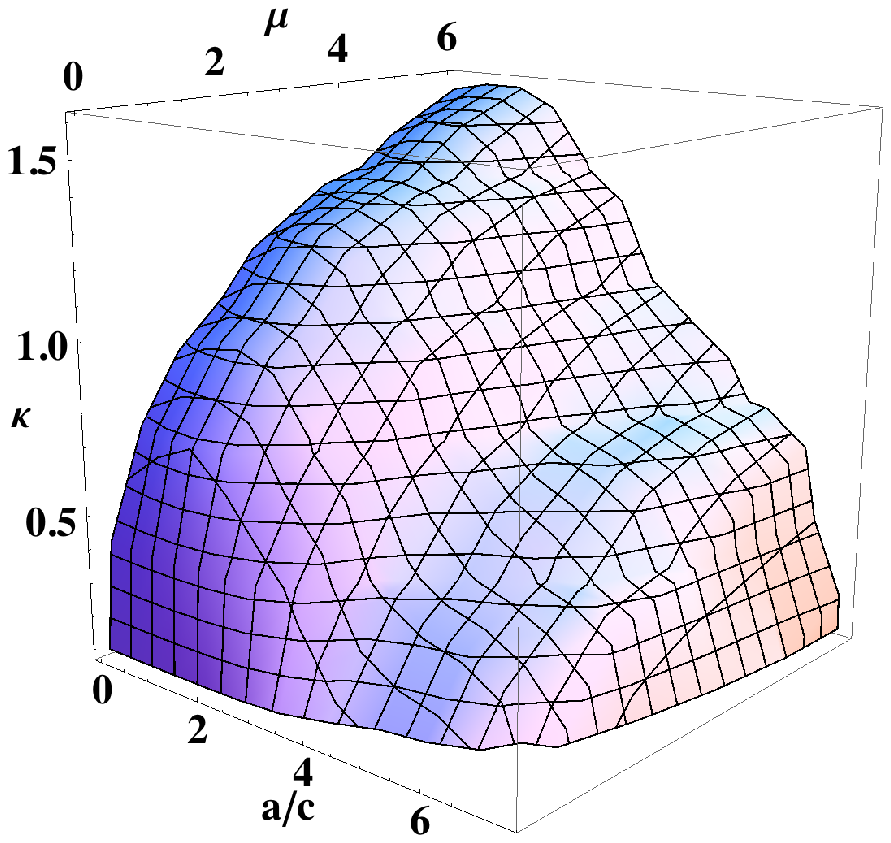}} \caption{Tricritical
surfaces in $(b/a,a/c,\kappa)$ with a fixed $\mu$ for the left
panel and $(\mu,a/c,\kappa)$ with a fixed $b/a$ for the right
panel, respectively, where vortices do not interact with each
other. The interaction potential is attractive inside the ellipse
while repulsive outside it.}\label{fig3}
\end{figure}

Next, we consider cases with a fixed $\kappa$. When $\kappa$ is
rather large, it is difficult to find the tricritical point
$\mu_{t}$, originating from smallness of the second term. In this
respect it is better to start from a small enough $\kappa$. Then,
the effective interaction is attractive when $\mu > \mu_{t}$ while
it becomes repulsive when $\mu < \mu_{t}$. Figure \ref{fig3} shows
a surface of tricritical points in the space of $(\kappa,\mu/T)$
with a fixed $T$, $a/c > 1$, and $b/a < 0$, where effective
interactions between vortices vanish exactly. The vortex
interaction is attractive inside the ellipse while it is repulsive
outside the ellipse. We claim that this ellipse serves a general
criterion for the fluctuation-driven first-order superconducting
transition in strongly coupled conformal field theories, possibly
occurring in the vicinity of quantum criticality.

In this study we try to answer how to classify strongly
interacting field theories, considering the nature of the
superconducting transition. The holographic superconductor model
is our main ansatz as an effective low energy theory, expected to
describe certain classes of strongly coupled conformal field
theories. The effective interaction between vortices is our
central object, allowing us to distinguish the type II
superconductor from type I, where the former will show the second
order transition while the latter will display the first order.
As shown, an asymptotic solution for a single vortex
configuration plays an essential role for the effective
interaction. The effective interaction between vortices turns out
to be a complicated function of both $\kappa$ and $\mu/T$, where the
parameter $\kappa$ is introduced to play basically the same role
as the Ginzburg-Landau parameter. We find a surface of tricritical
points in the parameter space of $(\kappa,\mu/T)$, where the
effective interaction vanishes, which separates the first order
from the second order, proposed to be a general criterion in
classifying quantum critical metals.

There are various unsolved questions in this direction. First of
all, a possible topological term such as the axion term
\cite{Axion_Vortex} may play an important role in the vortex
interaction. It can assign the U(1) charge to a vortex, modifying
their interactions. We suspect the possibility of the BKT
transition \cite{BKT}, resulting from their Coulomb interactions
due to the assigned U(1) charge, where $1/q^{2}$ in the momentum
space becomes $\ln r$ in two space dimensions. In addition to this
problem, the role of the pairing symmetry is not investigated,
where non $s$-wave superconductivity arises in strongly
interacting electrons \cite{QCP_Review1}. Furthermore, it should
be studied the role of fermions in the vortex interaction.

We would like to thank T.Albash for providing us with details in his work. K.Kim would like to thank Ki-myeong Lee and Chanju Kim for
helpful discussions. K.-S. Kim was supported by the National
Research Foundation of Korea (NRF) grant funded by the Korea
government (MEST) (No. 2010-0074542). Y.Kim acknowledges the Max
Planck Society(MPG), the Korea Ministry of Education, Science, and
Technology(MEST), Gyeongsangbuk-Do and Pohang City for the support
of the Independent Junior Research Group at the Asia Pacific
Center for Theoretical Physics(APCTP). K.Kim was supported by
KRF-2007-313- C00150, WCU Grant No. R32-2008-000- 10130-0.

\begin{widetext}

\appendix

\section{Numerical analysis for Eq.~(5)}

We start with discussion about a uniform solution in the original
holographic superconductor model [S. A. Hartnoll, C. P. Herzog, G.
T. Horowitz, Phys. Rev. Lett. {\bf 101}, 031601(2008)], where
$Q_t^s(z)$ and $\eta^{s}(z)$ depend only on the $z$-coordinate. It
is straightforward to derive equations of motion from Eq.~(3)
\begin{eqnarray}
&& {\eta}'' + \left(\frac{f'}{f} -\frac{2}{z}\right){
\eta}' + \left(  \frac{2 }{z^2 f}+\frac{\,{ Q_t}^2}{
f^2}\right)  \eta = 0 ,\qquad {{Q}_t}''
-\frac{ \eta^2}{z^2 f} {Q}_{t} = 0 .
\end{eqnarray}
The regularity at horizon ($z=1$) gives the following conditions
\begin{eqnarray} &&Q_t(1)=0 ,\quad Q_t'(1)=v, \quad Q_t''(1)=-\frac{u^2
v}{3} ,\quad \\&&\eta(1)=u ,\quad \eta '(1)=\frac{2 }{3} u,\quad \eta
''(1)=\frac{1}{2} u \left(-\frac{8}{9}-\frac{v^2}{9}\right).
\end{eqnarray}
Using the above conditions, one can find solutions in terms of $u$
and $v$, based on the shooting method. Near the boundary ($z=0$),
the solutions behave such as
\begin{eqnarray}
&& \eta(z) \sim O_1(u,v) z + O_2(u,v) z^2 + \cdots ,\qquad Q_t(z)
\sim \mu(u,v) + \rho(u,v)z+ \cdots .
\end{eqnarray}
When we are considering operators with dimension 1 or 2, we should
constrain solutions with either $O_2=0$ or $O_1=0$, respectively.
Therefore $u$ and $v$ are not independent but related with each
other. As a result, the space of solutions becomes one
dimensional, allowing us to take ``$v$" as a parameter for the
solution of the holographic superconducting state. In other words,
$v$ controls either temperature or charge density of the system.
According to the AdS/CFT dictionary, the total charge is given by
$\mathcal{Q} \sim -\int d^2 x\, \alpha\, Q_t'(0)$. When $T =
\frac{3\alpha}{4\pi}$ is fixed, the charge density varies as a
function of $v$. If the charge density is fixed, temperature
changes as a function of $v$. One can say a similar statement for
the chemical potential, $ \alpha \, Q_t (0)$. Inserting the
uniform solution of $Q_t^s(z)$ and $\eta^{s}(z)$ into Eq.~(5), we can solve them numerically.

Equation (5) has two parameters of $\kappa_{1}$ and $\kappa_{2}$.
Performing the scaling as discussed in the manuscript, we obtain
the following regularity conditions near the horizon for $A(z)$,
$B(z)$ and $C(z)$,
\begin{eqnarray}
&&A(1)=a ,\quad A'(1)=a \left(\frac{2}{3}+\frac{2 \kappa ^2}{3}\right) ,\quad A''(1)=-\frac{4 a}{9}-\frac{b u v}{9}-\frac{a v^2}{18}+\frac{4 a \kappa ^2}{9}+\frac{2 a \kappa ^4}{9},\\
&&B(1)=0 ,\quad B'(1)=b ,\quad B''(1)=-\frac{2}{3} a u v+\frac{1}{3} b \left(-u^2+2 \kappa ^2\right) ,\\
&&C(1)=c ,\quad C'(1)=c
\left(\frac{1}{3}-\frac{u^2}{3}\right),\quad C''(1) =c
\left(-\frac{5}{18}+\frac{u^2}{3}+\frac{u^4}{18}\right).
\end{eqnarray}
It is straightforward to see the scaling symmetries in Eq.~(5).
Equation for $C(z)$ remains invariant after scaling as $c\, \tilde
C(z)$ with a parameter $c$. In this case $c = 1$ is allowed due to
the boundary condition in Eq.~(A7). Equations for $A(z)$ and
$B(z)$ also allow scaling, unchanged after $a \tilde A(z)$ and $a
\tilde B(z)$. Therefore $b/a$ and $\kappa$ are only relevant
parameters, governing equations for $A(z)$ and $B(z)$. As a
result one may regard the asymptotic solution of the single
vortex configuration as
\begin{eqnarray}
s= a \tilde A\Bigl(v,\kappa,\frac{b}{a},z\Bigr) K_{0}\bigl(\sqrt 2 \kappa \bar
r\bigr),\quad q_t = a \tilde B\Bigl(v,\kappa,\frac{b}{a},z\Bigr) K_0\bigl(\sqrt 2 \kappa
\bar r\bigr),\quad q_\theta =c \,\tilde C(v,z) K_1 (\bar r)~,
\end{eqnarray}
where $\bar r$ means a re-scaled coordinate with $\kappa_2$. We
emphasize arguments in each function.

\section{Derivation of the variation for the grand potential}

Inserting Eq.~(6) into Eq.~(3), we find the following linearized
equations
\begin{eqnarray} \label{linear eom in sup}
&& \mathcal{D}^2 \delta \eta - m^2 \delta\eta - \delta \eta
{Q^{v}_\mu}^2  -2 \eta^{v} Q^{v}_\mu \delta Q^\mu =0 ,
\\\nn&&\mathcal{D}_\mu \delta B^{\mu\nu} - 2 \eta^{v} \delta \eta Q^\nu
-  {\eta^{v}}^2 \delta Q^\nu= - \mathcal{D}_\mu \delta X^{\mu\nu} ,
\end{eqnarray}
proven to be valid near the center of a vortex. One can see that
this approximation is reasonable only when $\delta \eta$ and
$\delta Q_t$ are both larger than $\delta Q_i^2$ and smaller than
$\eta^{v}$ and $Q_t^{v}$. The boundary of a Wigner-Seitz cell also
satisfies these conditions. In this respect the linearized
equations are valid not only near a vortex but also the boundary
of the cell. This is a simple extension of the observation in L.
Kramer, Phys. Rev. B {\bf 3}, 3821 (1971).

Inserting the vortex solution [Eq.~(6) with Eq.~(8)] into the
effective gravity action [Eq.~(1)] and expanding the action to the second order, we obtain the following expression for
the change of the grand potential in a cell
\begin{eqnarray}\label{delta Omega in sup}
\delta \Omega = \frac{1}{T} \int_{\hat i} dz d^2 x \partial_\mu
\left\{\sqrt{-g}\left[ \delta\eta \mathcal{D}^\mu \Bigl(  \eta^{v} +
\frac{1}{2}\delta \eta \Bigr) + \delta Q_\nu \Bigl( B^{v\mu\nu}+
\frac{1}{2}\delta B^{\mu\nu}  \Bigr)  \right]\right\} ,
\end{eqnarray}
where Eqs. (B1) and (B2) are utilized. In this derivation we need
to worry about singular parts from $X_{\mu\nu}$. $\delta X_{\mu\nu}$
vanishes identically because the singularity appears completely
outside the cell $\hat i$. The only term that we have to concern
is $\delta B_{\mu\nu} X^{\mu\nu}$, however this turns out to vanish
when we are considering the configuration of $\delta B_{\mu\nu} =
0$ at the origin of the cell.

Changing Eq.~(\ref{delta Omega in sup}) into surface integrals, we
have three kinds of boundaries. The first is the boundary at the
horizon $(z=1)$ of the black hole and the second is that of the
AdS space $(z=0)$. The last is the boundary of the Wigner-Seitz
cell. The first contribution vanishes identically thanks to
regularity conditions at the horizon. For the $z=0$ boundary, the
contribution must be considered carefully. Actually, this
contribution could be important, when a distance between vortices
is comparable to a size of a vortex. However, we are taking the
dilute gas limit, thus the variation from the single vortex
solution will be concentrated on boundaries of Wigner-Seitz cells.

The surface integral for $z = 0$ is given as follows
\begin{eqnarray}
\delta \Omega_{z=0} = \frac{1}{T} \int_{\hat i}  d^2 x
\sqrt{-g}\left[ \delta\eta \nabla^z \Bigl(  \eta^{v} +
\frac{1}{2}\delta \eta \Bigr) + \delta Q_\nu \Bigl( B^{v\,z\nu}+
\frac{1}{2}\delta B^{z\nu}  \Bigr)  \right]_{z=0} .
\end{eqnarray}
In the dilute limit $\delta \eta$ and $\delta Q_\mu$ have nonzero
values only near the boundary of a cell. Thus, the integration
range is effectively small. As positions of vortices are far from
each other, this contribution almost vanishes and it is much
smaller than the third contribution given by the integration along
the $z$ direction at the boundary of Wigner-Seitz cells. This
dilute approximation serves the validity of our calculation.
Therefore, our correction of the grand potential is well
approximated as
\begin{eqnarray}
\delta \Omega \sim \frac{1}{T} \int_0^1 dz \oint_{\hat i} dl \hat
n_i \sqrt{-g}\left[ \delta\eta \nabla^i \Bigl(  \eta^{v} +
\frac{1}{2}\delta \eta \Bigr) + \delta Q_\nu \Bigl( B^{v\,i\nu}+
\frac{1}{2}\delta B^{i\nu}  \Bigr)  \right] ,
\end{eqnarray}
where $\hat n_i$ is a unit vector orthogonal to the boundary of
the Wigner-Seitz cell. This leads to Eq.~(9).

\section{Derivation of Eq.~(9)}

In this section we will derive the following formulae
\begin{eqnarray}
&&\sum_{i\neq 0} \oint dl \,\hat n \cdot K_0(|r-r_i|)\hat
\theta(\vec r - \vec r_i) \times  \nabla \times \Bigl(K_0(r)\hat
\theta  + \frac{1}{2} \sum_{j\neq 0,i} K_0(|r-r_j|) \hat
\theta(\vec r - \vec r_j)  \Bigr)= \pi \sum_{i\neq0}K_0(|\vec
r_i|) ,
\\&&\sum_{i\neq 0} \oint dl \,\hat n \cdot K_0(|r-r_i|)  \nabla \Bigl(K_0(r) + \frac{1}{2} \sum_{j\neq 0,i} K_0(|r-r_j|)   \Bigr)= -\pi \sum_{i\neq0}K_0(|\vec r_i|) .
\end{eqnarray}
For convenience, we define a fictitious coordinate $w$ and vector
fields, $h_i(\vec r) \equiv K_0(|r-r_i|) \hat w$, where $r_i$ is a
center of a lattice. Then, the above equations can be written as
follows
\begin{eqnarray}
&&F1\equiv\sum_{i\neq 0} \oint d \vec S \cdot \left\{\Bigl( h_0  +
\frac{1}{2} \sum_{j\neq 0,i} h_j \Bigr) \times  \Bigl( \nabla
\times  h_i \Bigr) \right\}= \pi \sum_{i\neq0} h_i \cdot \hat w ,
\label{f1}
\\&&F2\equiv\sum_{i\neq 0} \oint d\vec S \cdot h_i  \times \nabla \times \Bigl(h_0 + \frac{1}{2} \sum_{j\neq 0,i} h_j   \Bigr)= -\pi \sum_{i\neq0} h_i \cdot \hat w ~~\label{f2},
\end{eqnarray}
where $d\vec S$ is an area element orthogonal to the boundary
surface of a cell, i.e, $d\vec S = dl~ dw ~\hat n$.

Using the divergence theorem and taking integration by parts, one
can rearrange $F1$ into
\begin{eqnarray}
F1&=&\sum_{i\neq0} \int d^3 x \Biggl\{  \nabla \times \Bigl(h_0 +
\frac{1}{2} \sum_{j\neq 0,i} h_j \Bigl)\cdot (\nabla \times h_i) - \Bigl(h_0
+ \frac{1}{2} \sum_{j\neq 0,i} h_j \Bigl) \cdot (\nabla \times \nabla
\times h_i ) \Biggr\} \nn &=&\sum_{i\neq0} \int d^3 x \Biggl\{
\nabla \times \Bigl(h_0 + \frac{1}{2} \sum_{j\neq 0,i} h_j \Bigl)\cdot
(\nabla \times h_i) + h_i \cdot \Bigl[ 2\pi \delta^2(r) \hat w -
\nabla \times \nabla \times \Bigl( h_0 +
\frac{1}{2}\sum_{j\neq0,i}h_j \Bigr)\Bigr]
\Biggr\}\nonumber\\&=& 2 \pi \sum_{i\neq 0}h_i(0) +\sum_{i\neq 0}
\oint d \vec S \cdot \Biggl\{ h_i \times \nabla \times \Bigl( h_0 +
\frac{1}{2}\sum_{j\neq0,i}h_j\Bigr) \Biggr\} ~~,
\end{eqnarray}
where we have used $\nabla \times \nabla \times h_i + h_i = 2\pi
\delta^2 (\vec r -\vec r_i) \hat w$ with some algebra. Actually,
the second term in the last is equal to $-F1$. In order to show
this, we consider the following combination,
\begin{eqnarray}\label{zzero}
&&\sum_{i\neq 0} \oint d \vec S \cdot \Biggl\{ \Bigl( h_0 +
\frac{1}{2}\sum_{j\neq0,i}h_j\Bigr)  \times \nabla \times h_i
\Biggr\} + \sum_{i\neq 0} \oint d \vec S \cdot \Biggl\{ h_i \times
\nabla \times \Bigl( h_0 + \frac{1}{2}\sum_{j\neq0,i}h_j\Bigr)
\Biggr\} \nn
&&\quad =\oint d\vec S \cdot \nabla \sum_{i\neq0} h_0 \cdot
\Bigl( h_0 +
\frac{1}{2}\sum_{j\neq0,i}h_j\Bigr) \nonumber\\
&&\quad = \oint d \vec S \cdot \nabla \Biggl\{  h_k
\cdot h_0 +(h_0 + h_k) \cdot \sum_{i\neq0,k} h_i+ \frac{1}{2}
\sum_{i\neq0,k} h_i \cdot\! \sum_{j\neq0,k,i} h_j \Biggr\} ,
\end{eqnarray}
where $k$ means one of other vortices. The integrand is symmetric
under interchange of $h_0$ and $h_k$. Now, we may take $\vec r_k$
as one of the nearest neighbor. Then, the symmetry means that the
integrand is a vector whose direction is along the boundary
surface of the cell. Thus, the above combination should vanish,
leading us to conclude that $F1 = \pi \sum_{i \neq0} h_i(0)\cdot
\hat w$. This argument is in parallel with that in L. Kramer,
Phys. Rev. B {\bf 3}, 3821 (1971). For $F2$, one can see that $F2
= -F1$ from Eq.~(\ref{zzero}). This completes our derivation.

\end{widetext}

\end{document}